\documentclass[a4paper]{jpconf}

\usepackage{graphicx}
\usepackage{amsmath, amssymb}
\usepackage{subfigure}
\graphicspath{{figs/}}

\usepackage{ulem}

\begin{document}

\title{A model for stable interfacial crack growth}

\author{Knut S Gjerden$^1$, Arne Stormo$^1$, Alex Hansen$^1$}

\address{$^1$ Department of Physics, Norwegian University of Science and 
Technology, N-7491 Trondheim, Norway}

\ead{knut.skogstrand.gjerden@gmail.com,arne.stormo@gmail.com,Alex.Hansen@ntnu.no}

\begin{abstract}
We present a model for stable crack growth in a constrained geometry. The 
morphology of such cracks show scaling properties consistent with self 
affinity. Recent experiments show that there are two distinct self-affine 
regimes, one on small scales whereas the other at large scales. It 
is believed that two different physical mechanisms are responsible 
for this. The model we introduce aims to investigate the two mechanisms 
in a single system. We do find two distinct scaling regimes in the model.
\end{abstract}

\section{Introduction}
\begin{figure}[ht]
 \begin{center}
  \includegraphics[width=2.8in]{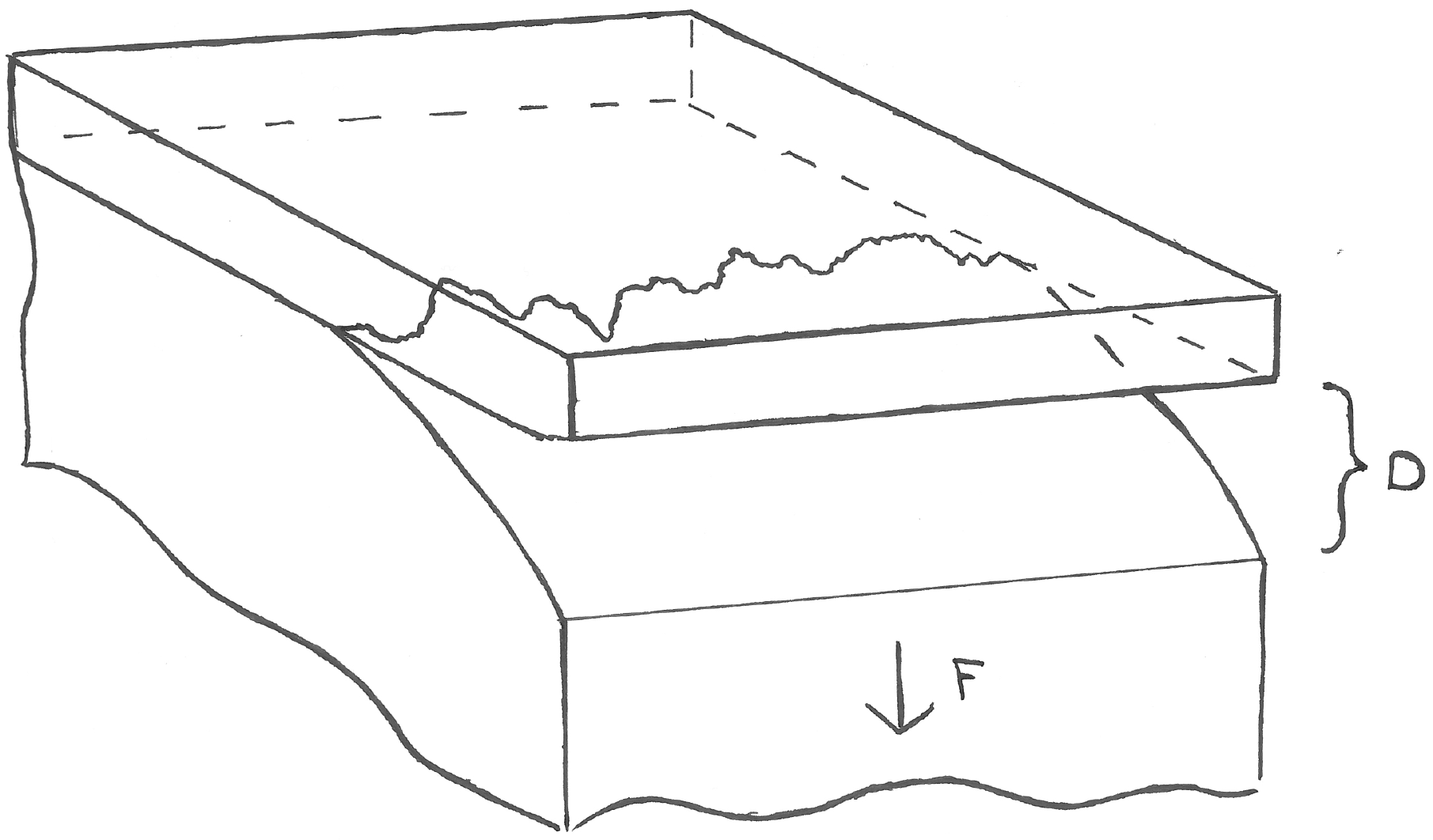}
  \caption{Sketch of the system we model. The two external 
control parameters are $F$ and $D$, representing 
either force-controlled or displacement-controlled loading, respectively.}
  \label{fig1}
 \end{center}
\end{figure}

On a fine enough scale, fractures are rough. How to describe this roughness 
quantitatively? An answer to this question was first proposed by Mandelbrot 
et al.\ in 1984 \cite{mandelbrot}: Fractures are fractal and their morphology 
may be described quantitatively  by a fractal dimension.  In 1990 Bouchaud 
et al.\ \cite{bouchaud1, bouchaud2},  using the more precise concept of 
self-affinity rather than fractality to characterize the fracture morphology 
of metals, proposed that the scaling properties are {\it universal,\/} i.e., 
they do not depend on the material that fractures.  An important conceptual 
step forwards was taken by Bouchaud et al.\ \cite{bouchaudjp} when they 
regarded the fracture surface to represent a ``footprint" of a passing 
fluctuating line, the crack front.  Schmittbuhl et al.\  
\cite{fluctuatingline} then made the suggestion that one may simplify 
the problem by constraining the crack growth to appear between two 
sintered plates and then follow the fluctuations of the crack front 
as it moves when the two plates are plied apart as show in Figure \ref{fig1}.  
In 1997, Schmittbuhl and M{\aa}l{\o}y \cite{maloy} realized 
this system experimentally by sintering two sandblasted plexiglass 
plates together.  

Schmittbuhl et al.\ \cite{fluctuatingline} presented a numerical model 
for constrained crack growth based on the fluctuating line picture.  
Here an elastic line moves through a plane covered with pinning centers.  
The elastic forces that the line feels is transmitted through 
the media constituting the two sintered plates.  Their main result 
was to determine the roughness exponent of the self-affine fracture front.  
If the front is given by $h=h(x)$, where the $x$-direction is orthogonal 
to the crack growth direction, then self affinity manifests 
itself through the following invariance: If $p(h,x)$ is the 
probability density that the crack is at height $h$ at $x$ when 
it is $h=0$ at $x=0$, then we have
\begin{equation}
\label{sadef}
\lambda^\zeta\ p(\lambda^\zeta h,\lambda x)= p(h,x)\;,
\end{equation}
where $\zeta$ is the {\it Hurst\/} or {\it roughness\/} exponent.  
Schmittbuhl et al.\ \cite{fluctuatingline} found $\zeta=0.35$.  
This value was later refined to $\zeta=0.39$ by Rosso and Krauth 
\cite{rosso}. The experiments \cite{maloy}, on the other hand, gave a much 
larger roughness exponent, $\zeta=0.55\pm0.05$.  In 2003, Schmittbuhl et 
al.\ \cite{coalescence} proposed a model based on the crack front 
propagating due to coalescence of damage appearing in front of it.  
This model gave a roughness exponent $\zeta=0.60\pm0.05$.  Recently, 
Santucci et al.\ \cite{santucci} have analyzed larger experimental 
systems than had been considered earlier.  They find a crossover 
between two scaling regimes in their data; a small-scale regime 
where $\zeta_1=0.60\pm0.05$ and a large-scale regime where 
$\zeta_2=0.35\pm0.05$.  Combining this result with the previous 
mechanism that has been suggested --- a fluctuating elastic line --- 
it is tempting to propose 
that the fracture coalescence mechanisms is at work on small scales, 
whereas on larger scales, the system effectively behaves as the 
fluctuating line model indicates.

It is the aim of this project to construct a single model capable of 
capturing both mechanisms and then to study the crossover between them.  
In the next section, we present the model, which is a variant of the 
fiber bundle model \cite{batrouni, pradhan}. The aim of this paper is 
to present this model in detail, including the computational 
side, which will be covered in Section 3. We present the 
treadmill technique in Section 4, which allows us to maintain a stable 
crack growth. In Section 5 we present our results, two different roughness
exponents $\zeta_1=0.45$ and $\zeta_2=0.3$.  These values are low
in comparison with the experimental values, $\zeta_1=0.60$ and $\zeta_2=0.3$.
As we will argue, this is probably due to the limited  system sizes we so far 
have considered.

\section{The model}
Our model is based on an idea originally proposed by Batrouni 
et al.\ \cite{batrouni}, but with some significant alterations. 
The basis of the model is a variation of the fiber bundle model, in which 
linear elastic fibers stretch according to the difference between their 
local displacement $u_i$ and a global displacement $D$,
\begin{equation}\label{base}
f_i=-k_i(u_i-D).
\end{equation}
This relation contains the proportionality constant $k_i$, which 
defines the response of the fiber. $L\times L$ fibers are arranged in 
parallel, with each fiber connected to two elastic blocks.
Each block behaves linearly elastic and can have its own elastic constant, 
but for simplicity and without loss of generality, one of the regions 
is set infinitely stiff and the other one elastic with Young's modulus $E$. 
The response of the fibers are transmitted through the elastic region 
via the Green's function \cite{johnson},
\begin{subequations}\label{green}
\begin{align}
  u_i&= \sum_jG_{ij}f_i,\\
  G^{}_{ij}= 
\frac{1-\nu^2}{\pi E a^2}&\iint^{a/2}_{-a/2}\frac{dx'dy'}{|\vec{r}_i(x,y)
-\vec{r}_j(x',y')|}. 
\end{align}
\end{subequations}
where $\nu$ is the Poisson ratio and $a^2$ is the area the force of each 
fiber acts on, thus giving the discretization of the system. 
$\vec{r}_i-\vec{r}_j$ gives the distance between fiber $i$ and $j$. 
The two regions are pulled apart either by controlling an external 
force or the displacement between them, as shown in Figure \ref{fig1}. 
The fibers break when they are stretched beyond a given threshold 
distribution. In matrix notation, the problem can be re-written as
\begin{equation}\label{mateq}
(\mathbb{I} +\mathbb{KG})\vec{f}=\mathbb{K}\vec{D},
\end{equation}
where $\mathbb{K}$ is an $L^2\times L^2$ diagonal matrix where 
$\mathbb{K}_{ii}=k_i$. To simplify the problem we keep all 
constants equal to unity. This is to isolate the disorder in the system 
to the thresholds, thus avoiding the complications of simultaneously 
dealing with two quenched disorder distributions. Because of 
the  $1/r$-dependence in the Green's function (\ref{green}), there are 
long-reaching forces in the system and every fiber is connected to 
every other fiber through the Green's function, making $\mathbb{G}$ a 
dense $L^2\times L^2$ matrix. $\mathbb{I}$ is the identity matrix and 
we have chosen to give the displacement vector $\vec{D}$ and 
solve for the forces in the $L^2$ vector $\vec{f}$.

We implement our model on a square lattice using bi-periodic boundary 
conditions with respect to the transmission of the forces. Such boundary
conditions are necessary due to numerical effects that we detail in the
following. We solve the matrix equation (\ref{mateq}) and locate the 
fiber with the highest ratio of strain to threshold. 
Then we break that fiber $i$ by setting $k_i$ to zero, 
indicating that it no longer has any load-bearing capability. 
The process is repeated until all fibers are broken.

\subsection{Loading schemes}
One of the new concepts of our model is the way we load our system. 
Providing a uniform displacement vector amounts to pulling the 
system apart, keeping the two surfaces parallel. 
Our main goal using this model was to observe a fracture front, 
meaning that we need some sort of gradient in the system to emulate 
the loading conditions of the experiments of Schmittbuhl and M{\aa}l{\o}y 
\cite{maloy}. There are at least three ways this can be done. 
The gradient can either be in the force, the displacement, or the thresholds.

Implementing a gradient in the force applied to the system would amount to 
solving the inverse problem since the system now would be load 
controlled. This is only possible by either reformulating the problem or 
adding a matrix inversion per broken fiber in a simulation, 
which would be very computationally costly. Given a choice 
between the displacement and the thresholds, choosing the 
thresholds enables us compare our results to that of gradient percolation
which then would be an extreme limit of the system where it is the 
treshold distribution that completely dominate how the breakdown process
proceeds rather than the force distribution of the fibers. Implementing 
a linear gradient in the thresholds models the pulling apart of the two blocks
at a constant contact angle.  This simplifies the problem and maintains
a close approximation to realistic loading.

\section{Numerical procedures}
We use the conjugate gradient (CG) algorithm with Fourier acceleration 
\cite{batrouni,FA} of 
the matrix multiplications to solve the system, meaning that the 
complete system we solve is 
\begin{equation}
(\mathbb{I}+\mathbb{KF}^{-1}\mathbb{F}^1\mathbb{G})
\mathbb{F}^{-1}\mathbb{F}^1\vec{f}=\mathbb{K}\vec{D}.
\label{compeq}
\end{equation}
The choice of an iterative solver is due to the matrix 
$\mathbb{KG}$ evolving from dense to sparse as the 
simulation progresses. We can Fourier accelerate because the 
Green's function, which in essence is a two-point 
correlation function, is only dependent on the distance 
between the points in question. This property makes the 
matrix $\mathbb{G}$ diagonal in Fourier space. 
Exploiting this, along with the symmetric properties of the 
resulting matrix on the left-hand side of (\ref{compeq}), 
all matrix multiplications are performed in 
Fourier space and we only store matrices of 
size $n=L\times L$, not $L^2\times L^2$, thereby 
avoiding large constraints on memory.

The higher the ratio between the Young's modulus and system size, i.e., 
the stiffer the system, the less impact the Green's function (\ref{green}) 
has on the system and the more well-behaved the equations are, 
leading to convergence in fewer iterations. For stiff systems, 
the pre-conditioning scheme of Batrouni et al.\ \cite{batrouni} 
works very well to further reduce the calculation times. Unfortunately, for 
softer systems, the Green's function can lead to large differences in 
neighboring forces which in turn leads to a very complicated energy 
landscape for the CG algorithm to traverse. This drastically increases 
the number of iterations required for convergence. In extreme cases, 
it does not converge at all. This also destroys the effect of the 
pre-conditioner, and we have not yet found an effective replacement. 

Another issue to address is the singularity in the 
Green's function (\ref{green}). To work around this, we use the solution 
of Love \cite{love}, expressing the displacement at point 
$i$ with coordinates $(x,y)$ due to a force $f_j$ acting on the area $a^2=4\varepsilon^2$ 
at the origin as
\begin{equation}
\label{eqlove}
\begin{split}
u_i= \frac{f_j(1-\nu^2)}{\pi E\varepsilon^2} \times \Bigg\{ 
\phantom{.}& 
\left(x+\varepsilon\right)\ln\left[\frac{\left(y+\varepsilon\right)+ \{\left(y+\varepsilon\right)^2
    + \left(x+\varepsilon\right)^2\}^{1/2} }{\left(y-\varepsilon\right)+
    \{\left(y-\varepsilon\right)^2 + \left(x+\varepsilon\right)^2\}^{1/2} }\right] \\
+& \left(y+\varepsilon\right)\ln\left[\frac{\left(x+\varepsilon\right)+
    \{\left(y+\varepsilon\right)^2 + \left(x+\varepsilon\right)^2\}^{1/2}
  }{\left(x-\varepsilon\right)+ \{\left(y+\varepsilon\right)^2 +
    \left(x-\varepsilon\right)^2\}^{1/2} }\right] \\
+& \left(x-\varepsilon\right)\ln\left[\frac{\left(y-\varepsilon\right)+
    \{\left(y-\varepsilon\right)^2 + \left(x-\varepsilon\right)^2\}^{1/2}
  }{\left(y+\varepsilon\right)+ \{\left(y+\varepsilon\right)^2 +
    \left(x-\varepsilon\right)^2\}^{1/2} }\right] \\
+& \left(y-\varepsilon\right)\ln\left[\frac{\left(x-\varepsilon\right)+
    \{\left(y-\varepsilon\right)^2 + \left(x-\varepsilon\right)^2\}^{1/2}
  }{\left(x+\varepsilon\right)+ \{\left(y-\varepsilon\right)^2 +
    \left(x+\varepsilon\right)^2\}^{1/2} }\right] \Bigg\}.
\end{split}
\end{equation}
Part of the derivation behind (\ref{eqlove}) relies on the 
assumption that the transition from the area uniformly acted on 
by the force $f_j$ to the area outside can be approximated as 
continuos. For stiff systems, this approximation is valid because there 
is little difference in the force carried by neighboring fibers. 
This remains valid over a large range of elastic moduli. For 
very soft systems, however, the difference in displacement and 
force carried can vary immensely across neighboring 
fibers, producing large local gradients eventually 
causing the assumption on a smooth transition to break down. 
We must emphasize that this is only in the case of the lower 
limit of the elastic modulus parameter, just as the case for 
the upper limit, an infinitely stiff system, equation 
(\ref{base}) reduces to
\[
f=kD,
\]
the purely global load sharing fiber bundle model. The point 
we are making is that when considering the Young's modulus $E$ 
as control parameter, the limits are one-sided in that the 
fundament for the equations breaks down before the lower limit of 
$E=0$ is reached. Between these two limits we show in this 
paper the existence of two distinct regimes controlling the roughness of the
crack front. 

To match the experimental setup as closely as possible, we want to 
use boundary conditions periodic along the emerging crack front. 
However, for soft systems, it is necessary to implement 
bi-periodic boundary conditions to aid convergence in the CG solver. 
In the next section we explain the treadmill technique, allowing us to follow 
the evolution of the crack front indefinitely.   
This also allows us to keep the crack front as far away from the edges as 
possible. For stiff systems, periodicity in one direction is sufficient 
and the priority is to have as much of unbroken material ahead of the 
front as possible. For soft systems, the Green's function becomes more 
important, forcing us to change to bi-periodicity and keep the front 
in the middle of system, as the crack front now runs the risk of being 
impeded by the mirror image behind it.

\section{The treadmill technique}
The basis of the treadmill technique is 1) the data is stored in matrix, 
or vector representation of a matrix, form and 2) as the simulation 
progresses, matrix elements are irreversibly changed to a common value. 
In our case, these requirements are automatically met, since our 
simulations are driven forward by breaking links, represented by the 
value of the load-bearing capacity of the element in question being 
set to zero. When the problem includes a gradient, the matrix changes 
from completely dense to sparse to completely empty in a logical manner. 
If the gradient is very sharp, half-way through the simulation, only the 
top half of the matrix will have non-zero elements and the other 
half will be computationally ``useless.'' Generally, only about half of 
the simulation can be used to gather reliable data, because the front 
needs time to fully develop in the beginning, and towards the end, it 
will eventually begin to feel the boundary conditions. To be 
conservative, on average, 40\% of a simulation is useable with 
respect to gathering trustworthy statistics on the crack front. 
Since models of this type are computationally costly and scales badly, 
increasing this percentage is a priority.

To increase the yield, of a given sample, the idea is to extend the 
middle part of a simulation.  Accomplishing this would amount to 
getting 100\% simulation yield after a simulation length of one 
system size has been reached. To do this, we have a criterion 
that tells us when the front has completely moved a 
pre-specified distance into the sample, and when this 
occurs, we shift all the values in the matrix downwards, effectively 
forgetting about the lowest completely broken part of the sample, and 
generate new, statistically identical material at the top of the 
matrix. Thus, we can control at what level the average position of 
the front should stay at, and follow this system 
indefinitely: A treadmill for interfacial fracture propagation.

\section{Results}
\begin{figure}[ht]
\begin{center}
  \subfigure{\includegraphics[width=1.8in]{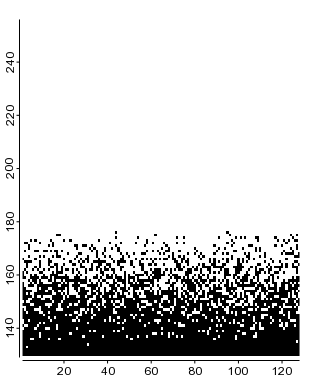}}
  \subfigure{\includegraphics[width=1.8in]{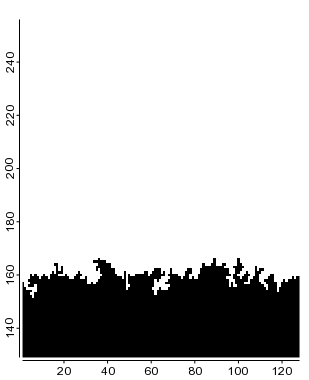}} \\
  \subfigure{\includegraphics[width=1.8in]{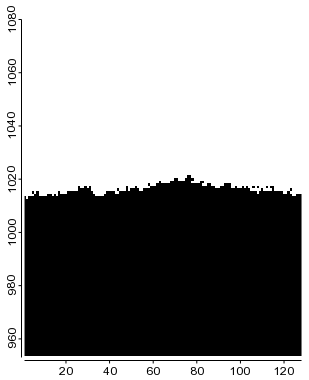}}
  \subfigure{\includegraphics[width=1.8in]{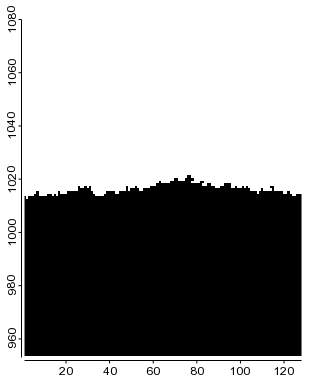}}
  \caption{Snapshots from the simulations showing clear 
differences between two distinct regions. The two upper 
images are from the same system, a system with a high Young's modulus, 
while the two lower images are from a simulation with a low Young's modulus. 
The leftmost images are direct printouts, whilst the rightmost 
images have been filtered to find the largest connected damage cluster, 
yielding the crack front. These images are the basis for a 
solid-on-solid analysis, to remove the overhangs. Note the 
difference in needed filtering between the two regions.}
  \label{figsim}
  \end{center}
\end{figure}

Our main result is to demonstrate that our model contains a transition 
from a regime where the crack front is primarily driven forward by 
coalescence of damaged regions ahead of the crack front, to a regime where 
the crack front appears to behave like an elastic line moving through 
the medium, by variation of the Young's modulus. 
These two types of behavior is exemplified in Figure \ref{figsim}. 
Note that, due to the treadmill technique, the crack fronts in the 
figure has propagated beyond the original size of the system, in this 
case $L=128$. The figure shows fronts from two simulations. 
In the first case, the system has a higher value for the Young's modulus, 
while the other example has a much lower value for $E$. In the stiffer 
system, we filter away the damage in front of and behind the crack front 
itself. This is illustrated in the upper part of 
Figure \ref{figsim}. As the system softens, the crack front changes 
behavior to not moving forward until all damaged material is part of 
the front, eliminating any ``damage islands'' as seen behind 
the front in the top left image in Figure \ref{figsim}. Tuning 
the system even softer, eventually only the crack front itself is 
seen as the interface between undamaged and damaged material, 
no isolated broken fibers ahead of the front is observed.

The choice of gradient in the system has an impact. In the results we 
present here, we have chosen a gradient that clearly demonstrates 
the different behaviors. A more complete study of the gradient effects will 
be presented elsewhere. For now, suffice to say that too steep a gradient 
will constrain the crack front to a smaller width, to the 
extreme limit where the crack front spans only one or two pixels. 
Setting a too low gradient will have the effect of increasing 
the width of the front to the point of the width of the 
crack front surpassing the length of the system. 
Zero gradient amounts to pulling the system apart, 
generating two parallel surfaces, but no crack front.

\begin{figure}[ht]
\begin{center}
\subfigure[Stiff system]{\includegraphics[width=2.8in]{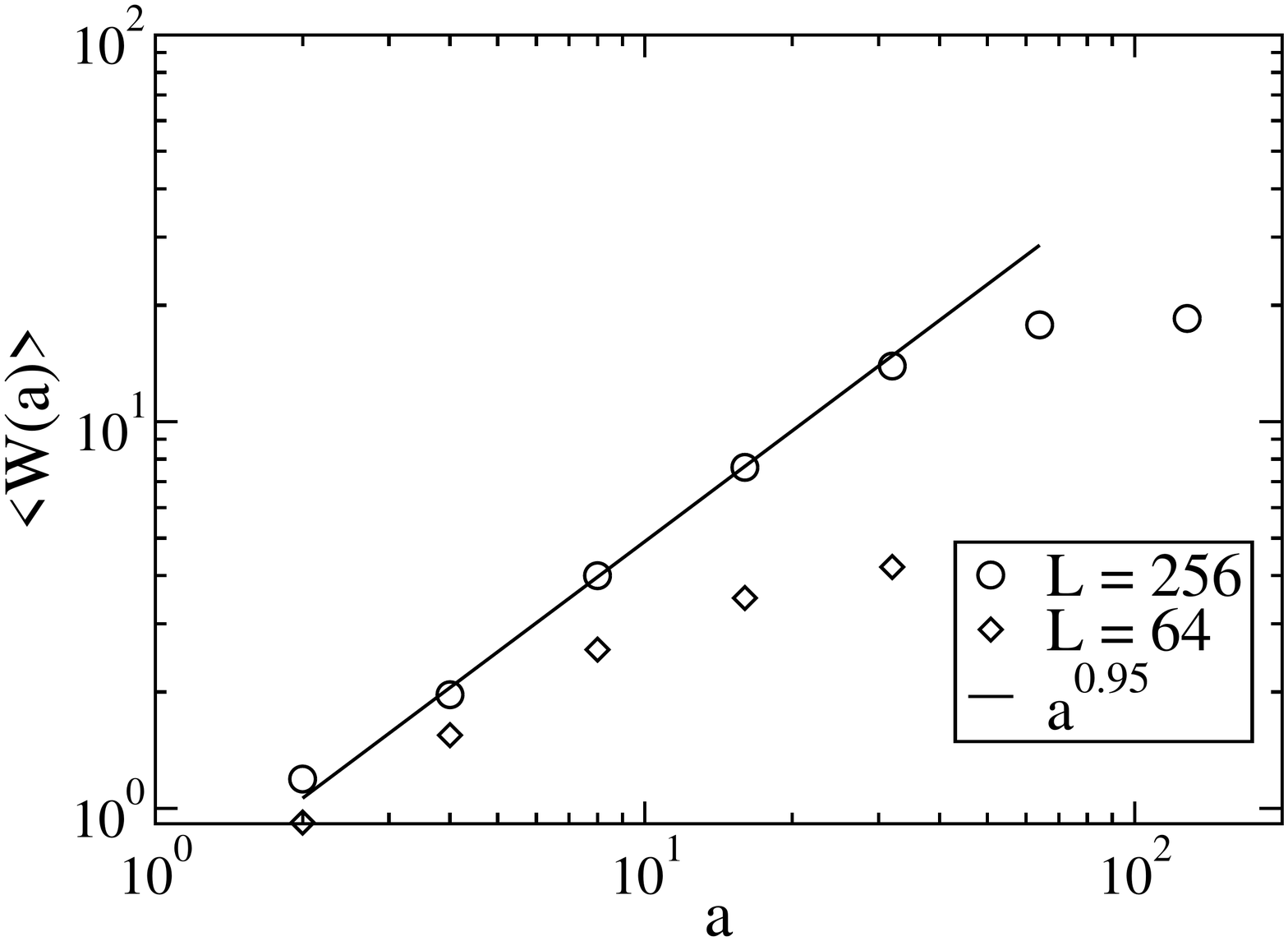}\label{hardlarge}}
\subfigure[Soft system]{\includegraphics[width=2.8in]{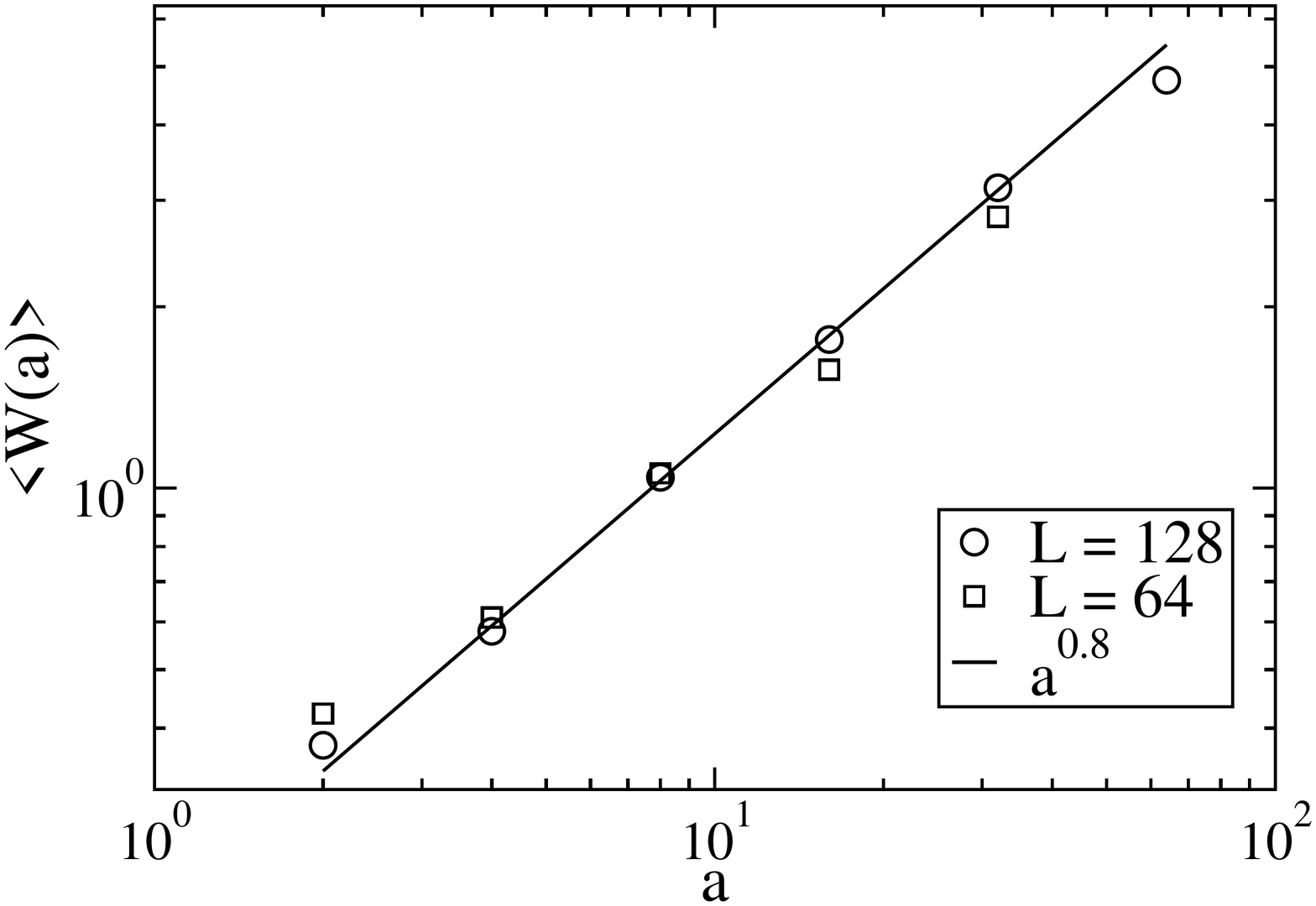}\label{softlarge}}
  \caption{Log-log plot of the average wavelet coefficients 
of a front extracted from the stiff region and from the soft region. 
Decent power laws are observed, indicating self-affine 
behavior of two different exponents. 
Two system sizes are plotted, indicating that the scaling exponents have
not yet reached their asymptotic values, so that our results are 
underestimations.}
  \label{large}
  \end{center}
\end{figure}

To analyze the front, we do a solid-on-solid (SOS) thus removing the overhangs.
Then we do an averaged wavelet coefficient (AWC) analysis 
\cite{wavmehrabi, wavsimonsen}, to determine the roughness exponent of the 
front. The AWC method consists of first doing a wavelet transform of the front
$h(x)$, and then, for each length scale $a$, averaging the resulting 
wavelet coefficients  $w(a,b)$ over position $b$ along the front. 
For self-affine signals, the averaged coefficients 
$W(a)=\langle|w(a,b)|\rangle_b$, should scale as
\begin{equation}
W(a)\sim a^{\zeta+1/2}.
\label{scale}
\end{equation}
The result of this analysis is shown in Figure \ref{large}, 
giving a roughness exponent of $$\zeta_1=0.45$$ for the stiff system in 
Figure \ref{hardlarge}, and $$\zeta_2=0.3$$ for the soft system in 
Figure \ref{softlarge}. We note that these values are lower than the values 
reported in the experiments \cite{santucci} and in the numerical studies,
$\zeta_{\textrm{c}}=0.6$ seen in the coalescence model \cite{coalescence} 
and $\zeta_{\textrm{l}}=0.39$ seen in the fluctuating line model \cite{rosso}. 
We also show the data for a smaller system size in the same figure.  We see
that those data would have given lower values for the roughness exponents.  
Hence, we believe that the largest system sizes we have presented here are still
too small for the effective roughness exponents to have attained their
asymptotic values.  We are currently testing 
several methods for running our code on massive 
parallel computer systems to increase the range of available system sizes.

If we do a small size analysis based on equations (\ref{base}) and 
(\ref{green}), and equation (\ref{eqlove}), using values from 
Schmittbuhl and M{\aa}l{\o}y \cite{maloy} where they found a response 
of plexiglass of $F\sim1$N at a displacement of 
$\delta=0.1$mm and a Young's modulus of $E\approx3.3$GPa, 
and assuming that the area covered by a single fiber is 
smaller than $\delta$, we find that the system sizes we have 
covered so far have room for forces acting at distances 
of the order of $0.1\sim1$mm. A system of $L=1024$ would 
probably reach on the order of 10mm, which would cover 
the range of the data used for the analysis of the 
experiment in both \cite{maloy} and \cite{santucci}. 
Based on the results of Santucci et al.\ \cite{santucci}, we believe that 
at larger system sizes we will find a crossover 
between the different scaling behaviors in the same simulation.

\section{Conclusions}
We have shown that our model contains a transition from one regime 
characterized by a scaling exponent of $\zeta_1=0.45$, to 
another regime characterized by a lower exponent of 
$\zeta_2=0.3$. This transition is achieved through the variation of a 
single parameter, the Young's modulus. We believe that these roughness 
exponents are slight underestimations of the exponents found in the 
coalescence model and fluctuating line model, 
respectively, making our model the first to capture 
these two fracture mechanisms simultaneously --- and that the idea that
the two roughness regimes seen in the experiments \cite{santucci} indeed
is correct.


\end{document}